\documentstyle[aps,12pt]{revtex}

\input epsf

\def\BNL {Department of Physics, Brookhaven National Laboratory, Upton, NY 11973\\}

\newcommand{\Dslash}{/ \!\!\!\! D}

\newcommand{\BE}{\begin{equation}}
\newcommand{\EE}{\end{equation}}
\newcommand{\BEA}{\begin{eqnarray}}
\newcommand{\EEA}{\end{eqnarray}}
\newcommand{\EL}{\nonumber\\}

\newcommand{\eg}{{\em e.\,g.\ }}
\newcommand{\etal}{{\em et al.\ }}

\newcommand{\ie}{{\em i.e.\ }}

\newcommand{\gbeta}{6/g^2}
\newcommand{\CS}{{\rm SU(N)_L\times SU(N)_R}}

\begin{document}

\title{ QCD with domain wall quarks } 
\author{T. Blum\thanks{email: tblum@penguin.phy.bnl.gov} 
and A. Soni\thanks{email: soni@penguin.phy.bnl.gov}}
\address{ \BNL }

\maketitle
\vskip .25in
\begin{abstract}
We present lattice calculations in QCD using a variant of
Kaplan fermions~\cite{KAPLAN,SHAMIR} which retain the 
continuum $\CS$ chiral symmetry
on the lattice in the limit of an infinite extra dimension. 
In particular, we show that the pion mass
and the four quark matrix element related to $K_0 - \bar K_0$
mixing have the expected
behavior in the chiral limit, even on lattices with modest extent
in the extra dimension, \eg $N_s=10$.
\end{abstract}

\vskip 0.5in
A fundamental property of the QCD Lagrangian is chiral symmetry, or
invariance under independent global flavor rotations of the left handed 
and right handed quark fields. Through the familiar Goldstone theorem,
the spontaneous breakdown of this symmetry
generates massless excitations, and hence chiral symmetry is
important to the dynamics of QCD. In reality the symmetry is also explicitly
broken by small quark masses, so the excitations are not massless, but
very light compared to masses of other hadrons.
Until now, lattice simulations of QCD have only accounted for this
symmetry approximately or not at all, relying on the continuum limit 
to restore it exactly. Since most simulations are far from the continuum limit,
this is not satisfactory.

The two popular discretizations of the Dirac action
used to date for lattice simulations of QCD are the Wilson quark action and the 
Kogut-Susskind action. For non-zero lattice spacing,
each breaks the continuum $\CS$ chiral symmetry.
The Wilson action includes a mass-like term, and the Lagrangian quark mass
renormalizes additively. In this case, the chiral symmetry breaking is 
severe, and fine tuning of the bare parameters is
required to obtain a massless pion, which defines the chiral limit. 
Indeed, these Goldstone bosons may not even be related to chiral symmetry 
in the nonperturbative regime,
but instead to the spontaneous breakdown of parity-flavor symmetry~\cite{AOKIETC}.
In addition, Wilson fermions lead to mixing of operators of differing chirality
which is absent
in the continuum. Physical results depend crucially on removal of these effects.
This has been a long-standing problem in lattice calculations 
of weak matrix elements with light hadrons~\cite{CBETAL,AS95} that use Wilson quarks.
Weak coupling perturbation theory is unable to accurately match the 
lattice operators to their continuum counterparts. Various methods have been
developed over the years to deal with this problem, and so far none is very
satisfactory~\cite{GAVELA,BS,GUPTAETAL,NPT}.

Kogut-Susskind fermions retain an exact ${\rm U(1)\times U(1)}$ remnant of
the continuum $\CS$ chiral symmetry\cite{KAWAMOTO}, so the bare quark mass 
renormalizes multiplicatively, and the chiral limit is obtained as in
the continuum, by taking the quark mass to zero. 
Because of this remnant chiral symmetry, Kogut-Susskind quarks 
are preferred for light hadron matrix element calculations and
simulations of QCD at non-zero temperature where it is now widely believed that
QCD undergoes a phase transition that restores chiral symmetry. This
phase transition happens at a critical temperature of roughly 150 MeV, 
or approximately the mass of the pion. Thus, the correct number of pions is important to 
describe the dynamics of the phase transition.
However, at finite lattice spacing the Kogut-Susskind action has only one Goldstone pion 
instead of the ${\rm N}^2-1$ in the continuum theory.

For these reasons it is desirable to use a discretization of the QCD action
that preserves chiral symmetry. Some time ago Kaplan~\cite{KAPLAN} proposed 
a discretization of the Dirac action for chiral fermions
that avoids the notorious fermion doubling problem
and hence could be used to formulate a nonperturbative chiral gauge theory. 
The impossibility of formulating such a theory that is local and gauge invariant
is summarized
by the Nielsen-Ninomiya no-go theorem~\cite{NN}. For each
chiral fermion that one starts with in the continuum,
a doubler of the opposite handedness appears that renders the lattice theory 
vector-like. This phenomenon is quite general and occurs already in
free field theory. However, for free Wilson-like fermions in d+1 dimensions, 
Kaplan showed that including a mass term with a defect along the 
extra dimension results in
a single chiral fermion which is bound to the d-dimensional defect. 
Thus a low energy effective theory of chiral fermions may be described
on the lower dimensional subspace of the defect. On a finite lattice with
periodic boundary conditions, an anti-defect appears with a bound chiral fermion
of opposite chirality, so again the theory is vector-like. However, if the
extent of the extra dimension is large, the fermions have exponentially 
small overlap and do not mix; one is free to study the lattice chiral fermion
on either wall. 
Sadly, when the chiral symmetry is gauged, the fermions mix, or if only
one is gauged, new particles appear in the spectrum, and
the chiral properties are probably lost~\cite{GOLTETAL}. But, see also the
overlap formalism of Ref.~\cite{NARNEU}.

The above considerations are irrelevant for QCD which is a vector gauge
theory, \ie right and left handed fermions couple to the gauge fields
with equal strength. The important point is that Kaplan's fermions 
retain the full $\CS$ continuum chiral symmetry on the lattice which makes them
an attractive alternative for simulating QCD.
Actually, at tree level one can show that 
the symmetry is broken explicitly by terms which are exponentially 
small in the size of the extra dimension~\cite{SHAMIR}. 
The main purpose of this work is to verify that these terms remain small in
nonperturbative simulations and do not spoil the chiral behavior of observables.

Below we describe calculations in QCD with a variant of Kaplan's 
original fermions that has been proposed by Shamir~\cite{SHAMIR,VRANASandJASTER}.
In this version only half of the lattice in the extra
dimension is used. The chiral fermions, one
left handed and one right handed, reside on opposite boundaries of the 
five dimensional space, and they interact with four dimensional gauge field
configurations
which are the same on each slice in the five dimensional world. The gauge
links in the fifth dimension are set to the identity except for the links
which connect the boundaries. Those links explicitly couple
the two chiral fermions with a strength equal to $-m$. In perturbation theory
it was shown in
Ref.~\cite{SHAMIR} that $m$ is proportional to the bare quark mass of the
vector fermion constructed from the left handed and right handed quarks 
on the opposite walls of the fifth dimension, $m_q=m M (2-M)$. Here $M$ is
the ordinary Dirac mass in lattice units for a five dimensional fermion.
Then in Ref~\cite{SHAMIRandFURMAN} it was shown that as the number of lattice
sites in the extra dimension, $N_s$, is sent to infinity,
the axial currents
of domain wall fermions satisfy nonperturbative Ward identities. 
Thus the chiral limit is
obtained by simply setting $m=0$, unlike in the case of 
regular Wilson fermions.

For details on Kaplan fermions and the variant of Shamir we refer the
reader to Refs.~\cite{KAPLAN,SHAMIR,SHAMIRandFURMAN};
here we simply state the quark action for one of the flavors~\cite{SHAMIRandFURMAN}.
\BEA
S_q &=&-\sum_{x,y,s,s^\prime}\bar\psi \left(\Dslash_{x,y}\delta_{s,s^\prime}+
\Dslash_{s,s^\prime}\delta_{x,y}\right) \psi, \EL
\Dslash_{x,y}&=& \frac{1}{2}\sum_{\mu}\left(
(1+\gamma_\mu)U_{x,\mu}\delta_{x+\hat\mu,y}+
(1-\gamma_\mu)U^\dagger_{y,\mu}\delta_{x-\hat\mu,y}\right)+
(M-4)\delta_{x,y}, \EL
\Dslash_{s,s^\prime}&=& \left\{ 
\begin{array}{ll}
\frac{(1+\gamma_5)}{2}\delta_{1,s^\prime}
-m\frac{(1-\gamma_5)}{2}\delta_{N_s-1,s^\prime}-\delta_{0,s^\prime} & s=0\EL
\frac{(1+\gamma_5)}{2}\delta_{s+1,s^\prime}+ 
\frac{(1-\gamma_5)}{2}\delta_{s-1,s^\prime}-\delta_{s,s^\prime} & 1\le s\le N_s-2\EL
-m\frac{(1+\gamma_5)}{2}\delta_{0,s^\prime}+ 
\frac{(1-\gamma_5)}{2}\delta_{N_s-2,s^\prime}-\delta_{N_s-1,s^\prime} & s=N_s-1
\end{array} \right . , \EL
\label{ACTION}
\EEA
where $\psi$ is a five dimensional Dirac fermion.
The $U_{x,\mu}$ are the usual four dimensional
gauge links and are elements of the color group SU(3). Note the Wilson term 
which 
removes the doublers is added to the action with minus the conventional sign. $x$ and
$y$ denote four dimensional coordinates while $s$ and $s^\prime$ denote coordinates along
the extra fifth dimension. $N_s$ is the number of sites in this dimension.
The simplest choice for operators which create and destroy light four dimensional quarks 
are~\cite{SHAMIRandFURMAN}
\BEA
q_x &=& \frac{(1+\gamma_5)}{2}\psi_{x,0}+ \frac{(1-\gamma_5)}{2}\psi_{x,N_s-1},\EL
\bar q_x &=& \bar\psi_{x,N_s-1}\frac{(1+\gamma_5)}{2}+ \bar\psi_{x,0}\frac{(1-\gamma_5)}{2}.
\label{quark ops}
\EEA
In fact, any operators localized near the boundaries that have finite overlap with the 
boundary states will do as interpolating operators for the quark states.

As a first step in implementing domain wall quarks for QCD, we 
measure observables constructed from the above operators
on existing configurations. The gauge field configurations that we 
have used are a set of 20 lattices generated 
with two flavors of Kogut-Susskind dynamical quarks
at gauge coupling $\gbeta=5.7$ and bare quark mass $m_{KS}=0.01$~\cite{COL}.
The lattice size for these configurations is $16^3\times 32$.
In all of our calculations, the Dirac mass $M$ was set to 1.7. 

For free quarks one must take $0<M<1$ to obtain massless states with 
strictly positive Green's functions~\cite{SHAMIR}. For $1<M<2$ there exists
a massless mode, but its Green's function oscillates like $(1-M)^{s}$.
This free field condition is renormalized at strong coupling as 
we found only heavy states for $M<1$. This can be seen from
perturbative arguments\cite{SHAMIR} and a simple mean
field treatment which gives $4-4U_0 < M < 6-4U_0$\cite{REFEREE}. Here $U_0$ is the fourth
root of the plaquette expectation value, or the mean field value of the
gauge link. The mean field propagator then
falls off like $(5-(M+4U_0))^s$. On the above configurations 
$U_0\approx 0.577$, yielding $M\approx 1.52$ for the most rapid decay 
of the propagator in the
extra dimension, and thus the smallest overlap of the boundary states. 
The above mean field result agrees fairly well with the ``optimal'' value of
$M=1.7$ which we determined by observing the exponential decay of the quark 
propagator in the extra dimension. As $M$ departs from this optimal value,
the extra dimension must be increased to avoid possible overlap of the
massless modes. While massless pions may still exist for $M<1$ at 
strong coupling, we did not observe them due to the relatively small 
extent of the lattice in the extra dimension. Thus, for efficient lattice 
simulations with domain wall quarks it is important
to determine the optimal value of $M$.
In this study we set the number of sites in the fifth direction 
to $N_s=4$ and 10 to
study the effects of finite size in the extra dimension on the chiral symmetry
properties of observables. Below we will see that for $M=1.7$, $N_s=10$ is
sufficiently large to suppress the overlap of the light modes.

\begin{figure}[hbt]
    \vbox{ \epsfxsize=6.0in \epsfbox[0 0 4096 4096]{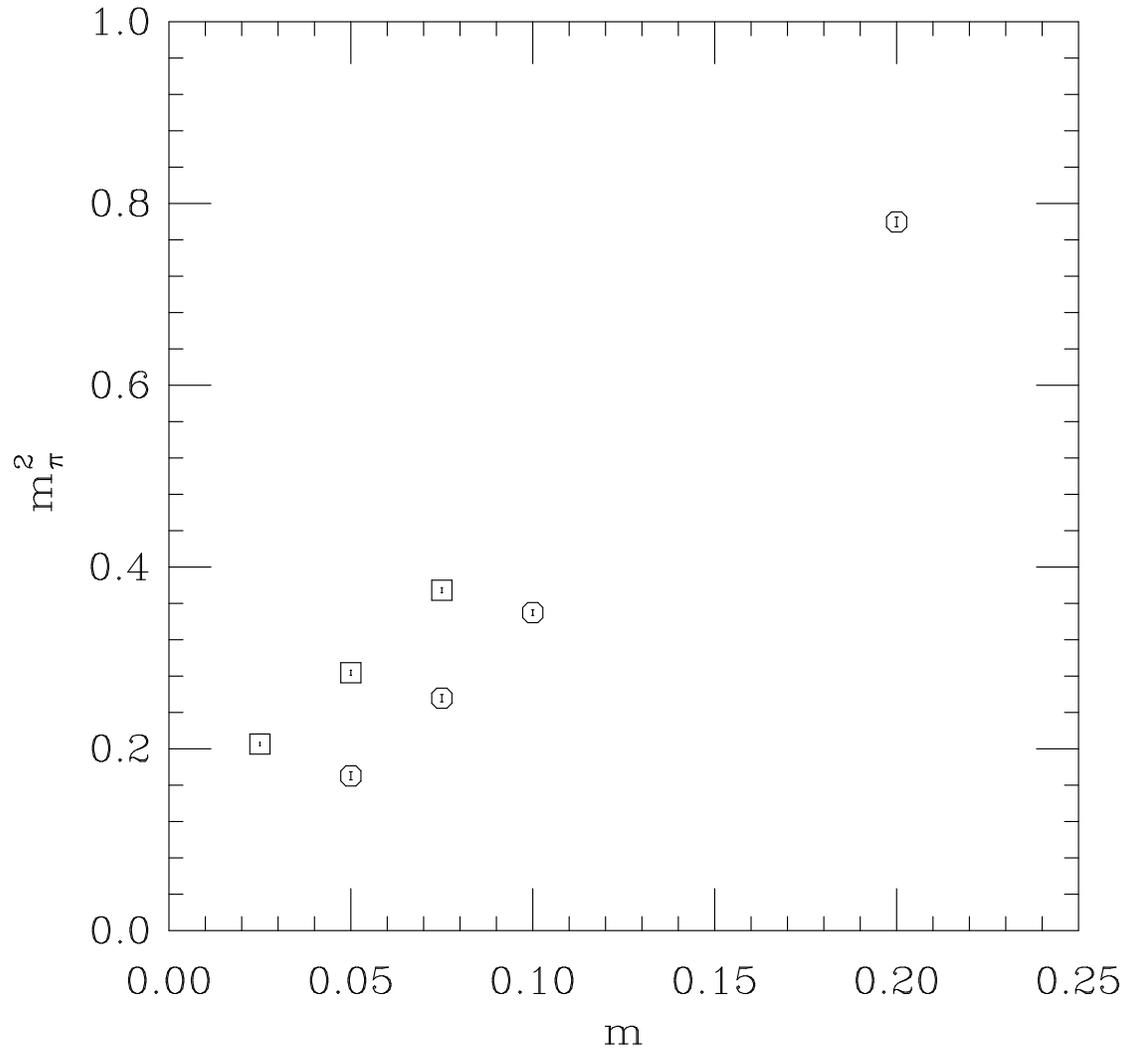} }
    \caption{The pion mass squared as function of $m$. $m$
	is proportional to the quark mass. The last two points for $N_s=10$
	(octagons) extrapolate linearly to zero well within statistical errors.
	The squares denote results for $N_s=4$.}
    \label{mpi squared}
\end{figure}

The simplest test of chiral symmetry (breaking) for domain wall fermions
is to measure the pseudo-scalar two-point correlation function and extract 
the mass. Chiral perturbation theory requires this mass to go to zero like
$\sqrt{m}$, or $m_\pi^2\propto m$. In Fig.~\ref{mpi squared} we show 
$m_\pi^2$ versus $m$ for the two values of $N_s$. The masses were extracted
from a single particle covariant fit to the pseudo-scalar correlator at large 
Euclidean time separation from the source (see below).
At $N_s=10$, the lightest two points linearly extrapolate to $0.0002 \pm 0.0160$ 
while for $N_s=4$ the results clearly miss the origin to the left. 
It appears that $N_s=10$ is large enough to suppress the explicit 
chiral symmetry breaking effects due to the overlap of the two chiral modes. 
The $N_s=4$ case shows that the approach to the $N_s\to\infty$
limit is from above which is expected since the Wilson term is 
added to the action with minus the conventional sign.

For a more stringent test of the chiral symmetry properties of domain 
wall quarks
we turn to the calculation of hadronic matrix elements, which as mentioned above
is problematic with conventional lattice operators.
The canonical example is the
$K_0-\bar K_0$ mixing matrix element, $M_{LL}$, required to determine the
$CP$ violating phase of the CKM matrix. Explicitly, $M_{LL}=\langle \bar K|
[\bar s \gamma_\nu ( 1- \gamma_5) d]^2|K\rangle$ and is expected to vanish as $m_K^2$
in the chiral limit. 
However, as we have mentioned, it is well known that the matrix element of this
operator constructed from Wilson quarks does not vanish in the chiral
limit. Only by fine tuning a linear combination of all 
four quark operators with
differing Dirac structures does one obtain the correct behavior. To date,
no satisfactory method of fine tuning has been found to solve this
problem~\cite{GAVELA,BS,GUPTAETAL,NPT},
so lattice operators which avoid this mixing are
much preferred, cf. Kogut-Susskind quarks and now domain wall quarks. 

\begin{figure}[hbt]
    \vbox{ \epsfxsize=6.0in \epsfbox[0 0 4096 4096]{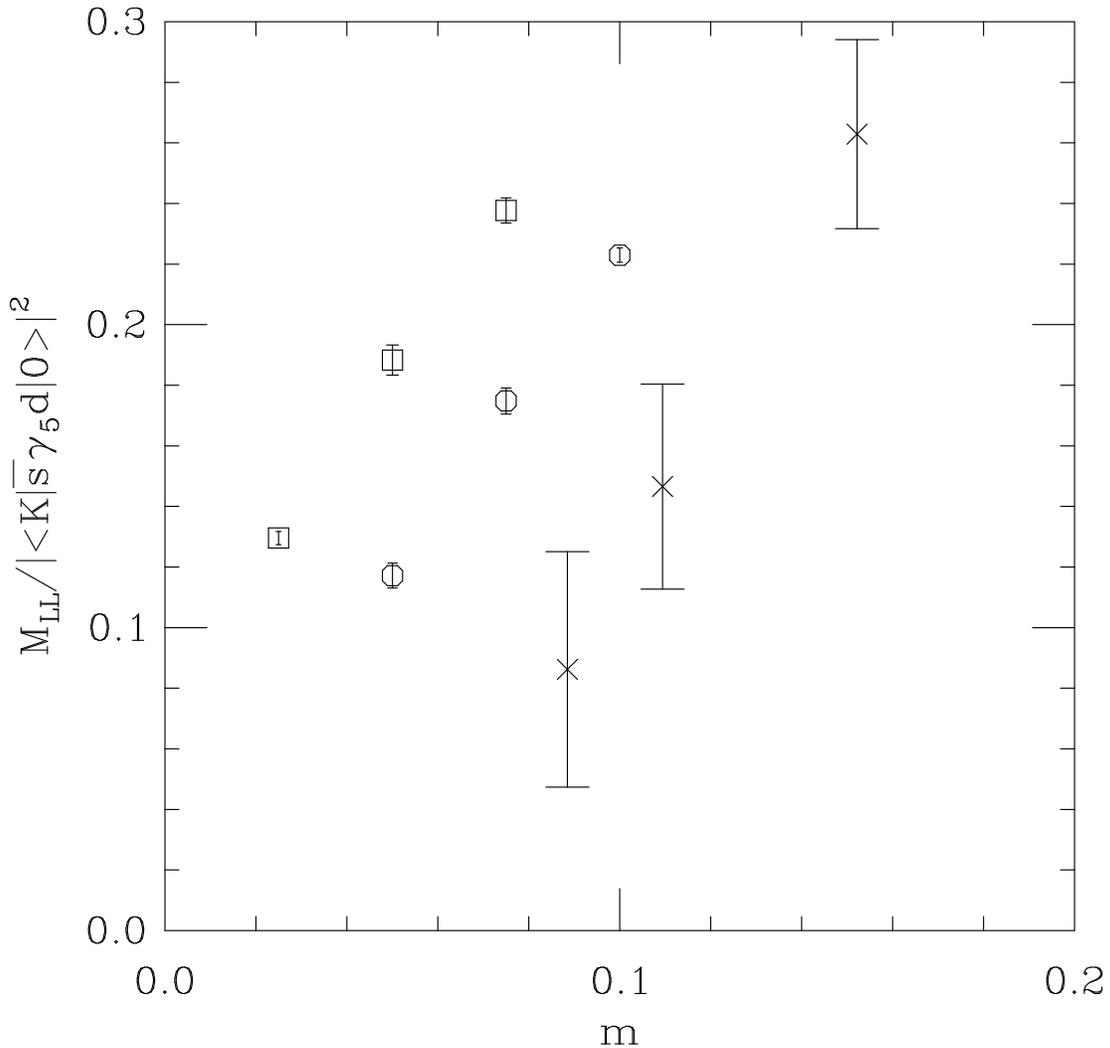} }
    \caption{The ratio of the four quark matrix element for $K_0-\bar K_0$
	mixing to the square of the pseudo-scalar density matrix element, 
	calculated with domain wall fermions (octagons($N_s=10$) and squares($N_s=4$)).
	The $N_s=10$ curve exhibits the correct behavior in the chiral limit. Also
	shown is the result using the same gauge field configurations 
	for Wilson quarks (crosses) which extrapolates to zero far from $m=0$ (note that
	for Wilson quarks the quark mass is defined as the difference of the inverse 
	quark hopping parameter with the inverse critical hopping parameter, 
	$m\equiv \frac{1}{2}(\kappa^{-1}-\kappa_c^{-1})$).}
    \label{matrix element ratio}
\end{figure}

In Fig.~\ref{matrix element ratio}
we show $\langle \bar K|[\bar s \gamma_\nu ( 1- \gamma_5) d]^2|K\rangle/
|\langle 0|\bar s\gamma_5 d|K\rangle|^2$ versus $m$
for both values of $N_s$ and the usual Wilson quark operator
whose mixing coefficients were determined using lattice weak coupling
perturbation theory~\cite{GM,BDS} with a boosted coupling~\cite{LM}.
This ratio has the same mass dependence 
as $M_{LL}$ in the chiral limit since
the denominator is just proportional to the square of the kaon decay constant. 
From Fig.~\ref{matrix element ratio} it is clear that
the $N_s=10$ result exhibits the required behavior in the
chiral limit: the values at the two smallest
quark masses extrapolate linearly to $.002\pm .015$ at $m=0$.
On the other hand, the $N_s=4$ domain wall quark and Wilson quark results
significantly miss the origin. We note that while the errors are considerably
larger for the Wilson case, the results at each quark mass are highly correlated
and a fit which accounts for this does not go near the origin. Indeed, previous
lattice calculations have shown that increased statistics do not remedy the problem.
Also note that the domain wall results again approach the large $N_s$ limit from above.

The ratios in Fig.~\ref{matrix element ratio} corresponding to domain wall
quarks are determined with the following procedure. First we calculate
five dimensional propagators from wall sources at times 0 and 31
(the ends of the lattice) on each boundary in the fifth dimension. 
The quark propagators are then constructed using Eq.~(\ref{quark ops}).
Because of the chirality projection operators in Eq.~(\ref{quark ops}),
only two source spins on each boundary are required.
We find that the added cost of inverting the five dimensional fermion matrix scales
linearly with $N_s$.
Next, these propagators were contracted using point sinks to form
three correlation functions on each gauge field configuration,
a two-point pseudo-scalar correlation function from each
wall source and a three-point correlation function from the same
two wall sources. Each correlation function is averaged over configurations.
The ratio is found by dividing the three-point correlation function by
the product of the pseudo-scalar
correlators, all three being evaluated at the same sink time slice.
For large time separations between the sink time slice
and the two wall sources, the result is a constant.
Our results were averaged over sink
time slices 10 to 20 where this constant plateau was evident.
The errors displayed in Fig.~(\ref{matrix element ratio}), which are statistical,
are calculated from a jackknife procedure.
In fact, simultaneous covariant fits
to all three correlation functions, which have good $\chi^2$,
give similar results. The Wilson quark results were calculated in similar fashion, but
with point sources.

A similar ratio as shown in Fig.~\ref{matrix element ratio}, but with
$\langle 0|\bar s \gamma_5 d|K\rangle$ replaced by 
$\langle 0|\bar s \gamma_5 \gamma_0 d|K\rangle$, gives 
the parameter $B_K$ which chiral perturbation theory 
predicts to be non-zero in the chiral limit. 
We are in the process of calculating this
important phenomenological quantity. Preliminary measurements indicate that the
ratio depends weakly on $m$ and is non-zero as $m\to 0$. 
The numerical value of $B_K$ is scale 
dependent (as is $M_{LL}$) and thus requires
a renormalization factor. In addition, finite lattice spacing
effects must be studied by simulating at several couplings
to understand the continuum limit of $B_K$. Thus much work needs
to be done before a meaningful value can be given.

The above results show that
domain wall quarks exhibit excellent chiral behavior, even at $g^2\sim 1$. 
In particular
the bare quark mass appears to receive only multiplicative renormalization,
\ie no fine tuning of the couplings is required to reach the chiral limit,
in contrast to the case with Wilson quarks. Strictly speaking this is true 
only in the limit $N_s\to \infty$; however, residual chiral symmetry 
breaking effects coming from finite $N_s$, which were not detectable in our calculations
with $N_s=10$, are in principle removable by simply increasing $N_s$.
These encouraging results also show that domain wall quarks
are a vast improvement 
over Wilson quarks for matrix element calculations. Again, there is no fine
tuning or cumbersome operator subtractions necessary to obtain the correct
chiral behavior. Similarly, they
are likely to prove useful for other QCD calculations where chiral symmetry
is crucial, \eg the QCD phase transition for non-zero temperature where current
simulations suffer from explicit chiral symmetry breaking effects.
Finally, we re-emphasize that domain wall quarks at $N_s=10$ retain the full
chiral symmetry of continuum QCD to a remarkable degree, unlike
either Wilson or Kogut-Susskind quarks.

We thank Y. Shamir and M. Creutz for suggesting the use of domain wall fermions 
for matrix element calculations and M. Creutz for helpful discussions.
Our domain wall fermion code relies heavily on the four dimensional MILC
code\cite{MILCCODE}, which we gratefully acknowledge.
This research was supported by US DOE grant
DE-AC0276CH0016. The numerical computations were carried out on the 
NERSC T3D and the SDSC Paragon.

\newpage

\end{document}